\documentclass[sigconf,screen]{acmart}

\usepackage{graphicx} 
\usepackage{stfloats} 
\usepackage{enumitem}
\usepackage{multirow}
\usepackage{longtable}
\usepackage[most]{tcolorbox}
\usepackage{tabu}
\usepackage{listings}
\usepackage{amsmath}
\usepackage{amsfonts}
\usepackage{mathtools}
\usepackage{amsthm}
\usepackage{threeparttable}
\usepackage{color}
\usepackage{array}
\usepackage{algorithm}  
\usepackage{algorithmicx}  
\usepackage{algpseudocode}
\usepackage{makecell}
\usepackage{subcaption}
\usepackage{caption}

\definecolor{lightgreen}{rgb}{0.25, 0.63, 0.4375}
\definecolor{darkblue}{rgb}{0.02, 0.16, 0.49}

\definecolor{darkgreen}{rgb}{0, 0.5, 0}
\definecolor{darkred}{rgb}{0.72,0.04,0.04}

\newcommand{\code}[1]{{\ttfamily #1}}

\author{Shangtong Cao}
\affiliation{%
 \institution{Beijing University of Posts and Telecommunications}
 \country{China}
 }

\author{Ningyu He}
\affiliation{%
 \institution{Peking University}
 \country{China}
}

\author{Xinyu She}
\affiliation{%
 \institution{Huazhong University of Science and Technology}
 \country{China}
}

\author{Yixuan Zhang}
\affiliation{%
 \institution{Peking University}
 \country{China}
}

\author{Mu Zhang}
\affiliation{%
 \institution{University of Utah}
 \country{America}
}

\author{Haoyu Wang*}
\affiliation{%
 \institution{Huazhong University of Science and Technology}
 \country{China}
}

\newcommand{\framework}{\textsc{WRTester}}

\setcopyright{none}
\renewcommand\footnotetextcopyrightpermission[1]{} 

\title{{\framework}: Differential Testing of WebAssembly Runtimes via Semantic-aware Binary Generation}

\begin{document}

\begin{abstract}
Wasm runtime is a fundamental component in the Wasm ecosystem, as it directly impacts whether Wasm applications can be executed as expected. Bugs in Wasm runtime bugs are frequently reported, thus our research community has made a few attempts to design automated testing frameworks for detecting bugs in Wasm runtimes. However, existing testing frameworks are limited by the quality of test cases, i.e., they face challenges of generating both semantic-rich and syntactic-correct Wasm binaries, thus complicated bugs cannot be triggered. In this work, we present {\framework}, a novel differential testing framework that can generated complicated Wasm test cases by disassembling and assembling of real-world Wasm binaries, which can trigger hidden inconsistencies among Wasm runtimes. For further pinpointing the root causes of unexpected behaviors, we design a runtime-agnostic root cause location method to accurately locate bugs. Extensive evaluation suggests that {\framework} outperforms SOTA techniques in terms of both efficiency and effectiveness. We have uncovered 33 unique bugs in popular Wasm runtimes, among which 25 have been confirmed.
\end{abstract}

\maketitle

\section{Introduction}

WebAssembly (Wasm), a low-level bytecode format, was proposed by several Internet giants in 2017~\cite{bring-web-with-wasm}. Due to its excellent portability, native-like speed, compacted size, and safety guarantee, Wasm is gaining growing popularity. In 2022, more than 97\% existing browsers have supported Wasm. Beyond web browsers, Wasm has been favoured in a wide range of domains, including mobile apps, blockchain, and IOT, etc.
Wasm can be regarded as the compilation target for almost all main-stream high-level programming languages, e.g., C, C++, Go, and Rust~\cite{c/c++,Rust,Go}. 
Wasm binaries are executed in Wasm runtimes, which is similar to a virtual machine that serves as an intermediate layer between the WASM binaries and the underlying system.
Currently, lots of Wasm runtimes have been implemented and actively maintained on GitHub, like Wasmtime~\cite{wasmtime}, Wasmer~\cite{wasmer}, and WasmEdge~\cite{wasmedge}.

Wasm runtimes, play a key role in the ecosystem, as it directly impacts whether Wasm applications can be executed as expected. However, a variety of Wasm runtime specific bugs have been reported from time to time.
For example, Zhang et al.~\cite{zhang} have empirically analyzed over 300 real-world bugs of Wasm runtimes and created a taxonomy of 31 bug categories of Wasm runtimes. 
Thus, some fellow researchers in our community proposed to develop automated methods for detecting bugs in Wasm runtimes. 
For example, Jiang et al.~\cite{wasmfuzzer} have uncovered that Wasm runtimes may cannot execute the given Wasm binary correctly by adopting coverage-guided fuzzing. WADIFF~\cite{wadiff} further adopted symbol execution to generate lots of Wasm binaries and conducted differential testing to identify implementation bugs of runtimes.

Although recent automated testing approaches have shown promising results in identifying Wasm runtime bugs via Wasm binary generation, they are however limited by their inability to generate semantically rich binaries (see \S\ref{sec:challenge}), thus complicated bugs cannot be triggered. 
For example, WADIFF can only test the implementation bugs at the single instruction level, as it cannot generate test cases with a large number of instructions, while real-world Wasm binaries are indeed much more complex. Specifically, there exist over 430 kinds of instructions and 13 types of sections with different functionalities in Wasm, indicating that Wasm is a bytecode format with rich semantics. For comprehensive testing, it is crucial and necessary to ensure that the generated Wasm binaries can cover as much semantics as possible. Further, it is impossible to generate Wasm binaries arbitrarily, as each Wasm binary should be validated for syntactic correctness before executing, including stack balance verification etc, which ensures that the generated Wasm binaries should be syntactic-correct.

\textbf{This work.}
We present {\framework}, a novel differential testing framework for Wasm runtimes that can generate syntactic-correct and semantic-rich Wasm binaries. To achieve this, we have designed a dedicated algorithm to extract basic elements from real-world Wasm binaries, and randomly assemble them into Wasm binaries with valid syntax and rich semantics. Based upon this, {\framework} further applies different levels of mutation strategies (e.g, AST-level and module-level mutation) on generated Wasm binaries to increase their diversity. {\framework} sends the generated Wasm binaries to Wasm runtimes simultaneously to investigate inconsistent behaviors. As different forms of input binaries that cause the same inconsistent state may be reported redundantly, which would result in an excessive amount of generated inputs. Thus, for further pinpointing the root causes of inconsistencies, we proposed a root cause identification algorithm that can accurately locate bugs on a function-level or even instruction-level.
Extensive experiments show the superiority of {\framework} over state-of-the-art techniques in both terms of efficiency and effectiveness, i.e., {\framework} can generate 6.0x and 148.8x Wasm binaries that can lead to inconsistencies over two baseline tools. By applying {\framework} to four representative Wasm runtimes, it has generated over 167K Wasm binaries that can lead to inconsistencies among Wasm runtimes, which are further located to 33 unique bugs. With our timely disclosure, 25 bugs has been confirmed by runtime developers, and 11 of them have been fixed with our aid, by the time of this writing.

We make the following main contributions in this work:
\begin{itemize}
	\item We propose {\framework}, a novel differential testing framework that can generate syntactic-correct and semantic-rich Wasm binaries by disassembling and assembling of real-world Wasm binaries, which significantly increases the diversity of generated Wasm binaries in terms of semantics.
 
	\item We design a runtime-agnostic root cause identification algorithm that can accurately pinpoint the location of bugs, which significantly eases the burden of runtime developers for further verification and bug patching.
 
	\item {\framework} have identified 33 unique bugs that can lead to unexpected behaviors for mainstream Wasm runtimes, among which 25 have been confirmed and 11 have been patched with our timely disclosure.
\end{itemize}

\section{Background \& Motivation}

\subsection{WebAssembly}
\label{sec:backgroud:wasm}
WebAssembly (Wasm) is an emerging stack-based binary format that can be compiled from mainstream high-level languages.
Except for the original four primary data types, i.e., \texttt{i32}, \texttt{i64}, \texttt{f32}, and \texttt{f64}, \texttt{v128} is introduced in the recent specification to support SIMD instructions~\cite{simd}.
Specifically, the leading \texttt{i}, \texttt{f}, and \texttt{v} respectively refer to \textit{integers}, \textit{float numbers}, and \textit{vectors}, while the following number stands for the length in bits. 
Each Wasm binary is composed of 13 sections~\cite{format}, and complex functionalities can only be achieved by coupling sections, e.g., implementing a function requires three sections. The \textit{type section} declares the function signatures, the \textit{code section} contains all local variables and the function body, and the \textit{function section} keeps a mapping relation from the type index to the function index.
Wasm applies a specific \textit{structured control flow}.
Instructions in Wasm are divided into \textit{code blocks}, which can be led by \texttt{block} or \texttt{loop}.
Code blocks can be nested, but the context of an inner code block is independent to its outer ones.
Moreover, there are no \texttt{goto}-like instructions in Wasm, indicating the control flow cannot be directed arbitrarily.
Instead, \texttt{br} and \texttt{br\_if} are used for control flow jumps, whose target has to be the heading (led by \texttt{loop}) or tailing (led by \texttt{block}) of its outer code block.

Each Wasm binary will be statically verified on its syntactic validity before executing.
The verification mainly focuses on the stack. On the one hand, it verifies if each instruction can take operands whose types match the ones defined in the specification. On the other hand, it verifies the stack balance, i.e., the behavior of each block and function is consistent with its signature.

\subsection{Wasm Runtime}
\label{sec:backgroud:runtime}
Wasm runtime provides an executing environment for Wasm binaries in various hardware and operating systems~\cite{zhang}. They play a vital role in supporting Wasm-based functionalities in blockchain platforms~\cite{eosio} and embedded devices~\cite{embed}, enabling the deployment of lightweight, high-performance applications in resource-sensitive environments.
Additionally, except for the inefficient \textit{interpreting} mode, both \textit{JIT (Just-In-Time)} and \textit{AOT (Ahead-Of-Time)} are adopted by some runtimes to improve the performance.
Wasm runtime is also responsible for handling interactions between Wasm binaries and the external environment. In the early stage, each runtime has its specific set of compiling tool chains and wrappers for APIs exposed by the operating systems, which results in a severe compatibility issue. Thus, WebAssembly System Interface (WASI)~\cite{wasi} emerges, which defines the function signatures of each API as well as its behavior. 
Currently, WASI is supported by lots of mainstream Wasm runtimes as well as their compiling tool chains.

\subsection{Motivation}
\label{sec:challenge}
Improper implementation of Wasm runtimes will significantly hamper the intentional design goal of Wasm, i.e., \textit{security} and \textit{efficiency}.
However, testing the correctness of the implementation of Wasm runtimes is challenging. Currently, only differential testing, one of the dynamic analysis methods, is adopted by existing studies~\cite{wadiff,zhang}.
This is because the static verification may be struggled with the complex logic in runtimes, and the pre-defined rules very likely import false positives.
Differential testing is a widely adopted technique that compares the outputs or states among different targets (i.e., different implementations of the same functionality) while giving an identical input. Moreover, it is independent of oracles, one of the main challenges faced by other dynamic analysis methods, like grey-box fuzzing.
Although differential testing seems to be a promising approach, there still exist some challenges for detecting bugs in Wasm runtimes, which can be summarized as follows:

\noindent
\textbf{Challenge \#1: Generating syntactic-correct and semantic-rich Wasm binaries}. 
As we mentioned in \S\ref{sec:backgroud:wasm}, each Wasm binary should be validated for syntactic correctness before executing.
Further, in Wasm, there exist over 430 instructions and 13 types of sections with different functionalities, indicating that Wasm is a bytecode format with rich semantics.
To reach the goal of comprehensive testing, instead of guaranteeing the syntactic correctness, it is also crucial and necessary to ensure that the generated Wasm binaries can cover as much semantics as possible.
In other words, \textit{the generated Wasm binaries should be syntactic-correct and semantic-rich}.
As for the syntactic correctness, the generated binaries should be stack-balanced, and the index reference among sections should be correct.
As for the semantic richness, on the one hand, the instructions in the generated Wasm binary should interact with as many sections as possible. On the other hand, these instructions should be covered as much as possible during execution at runtime.

\noindent
\textbf{Challenge \#2: Error locating to refine Wasm binaries according to root causes}.
During the differential testing process, when observing inconsistent results among different runtimes, it is challenging to pinpoint the root cause on each case.
Specifically, differential testing will generate lots of Wasm binaries, each of which is composed of at least hundreds of thousands of instructions. It is hard to locate which function or instruction indeed leads to a bug, even less pinpointing the root cause. Indeed, many inconsistent results observed during the testing process are introduced by the same bug.
However, implementing the bug diagnosing processes is challenging because not all Wasm runtimes come with developed debugging tools. Moreover, handling the compatibility issue among these runtimes also raises the concern of scalability.

\noindent
\textbf{Limitations of current tools}.
To the best of our knowledge, only two tools are available for differential testing Wasm runtimes, i.e., wasm-smith~\cite{wasm-smith} and WADIFF~\cite{wadiff}.
Specifically, wasm-smith is a Wasm test case generator proposed by the official community. It randomly selects instructions while considering the stack balance to guarantee the syntactic correctness.
As for WADIFF, it adopts symbolic execution to generate test cases for each instruction according to the specification.
However, both tools have certain limitations.

\begin{figure}[t]
  \begin{subfigure}[t]{0.4\columnwidth}
    \centering
    \includegraphics[width=\linewidth]{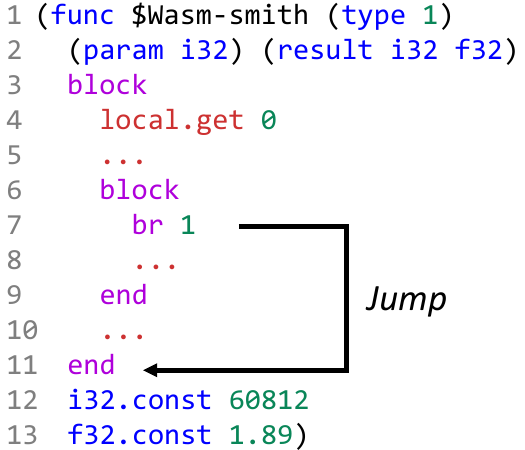}
    \caption{A Wasm binary generated by wasm-smith.}
    \label{fig:motivation:wasm-smith}
  \end{subfigure}
  \hfill
  \begin{subfigure}[t]{0.46\columnwidth}
    \centering
    \includegraphics[width=\linewidth]{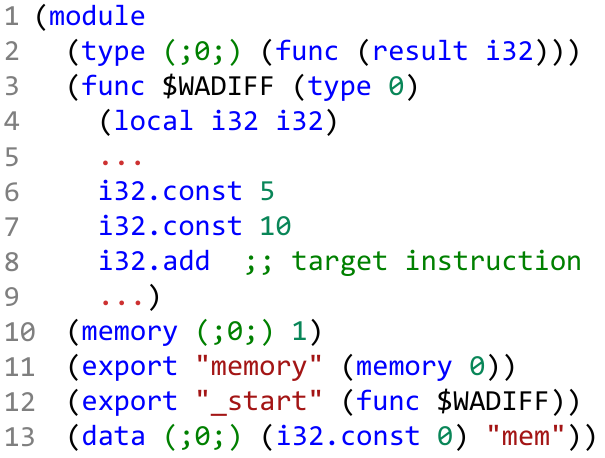}
    \caption{A Wasm binary generated by WADIFF.}
    \label{fig:motivation:wadiff}
  \end{subfigure}
  \caption{Comparison of Wasm binaries.}
  \label{fig:motivation:comparison}
\end{figure}

Specifically, wasm-smith ignores the semantics of the generated Wasm binaries, hindered by the \textbf{Challenge \#1}.
As shown in Figure~\ref{fig:motivation:wasm-smith},
the \code{br} instruction at L7 will direct the control flow to L11, i.e., the end of the function. In other words, instructions between L7 and L11 will be ignored by runtimes.
Additionally, L2 indicates that the results should be an \texttt{i32} and a \texttt{f32}. To ensure the stack balance, wasm-smith simply pushes two constant values (L12 and L13) of the corresponding types.
This means that this function can be aggressively optimized to only include the last two instructions.
Consequently, wasm-smith can only generate Wasm binaries composed of lots of meaningless instructions, which hampers both efficiency and effectiveness of differential testing.

As for WADIFF, which adopts symbolic execution on the Wasm specification of each instruction to generate test cases, it can only generate simple Wasm binaries ($\sim$10 -- 100 instructions) to verify runtimes.
As shown in Figure~\ref{fig:motivation:wadiff}, this Wasm binary verifies the implementation of \texttt{i32.add}.
WADIFF follows the Occam's Razor principle~\cite{Occams}, i.e., following the simplest control flow and giving only the necessary parameters generated by constraints.
In other word, it is hard for WADIFF to cover complex functionalities among 13 sections within a single Wasm binary.
Moreover, it can only generate a finite number of Wasm binaries after traversing all possible paths for each instruction, and its random mutation method is likely to generate invalid ones~\cite{wadiff}.
Last but not least, generating test cases against control instructions is not supported by WADIFF yet, e.g., \code{call\_indirect} and \code{loop}.
In summary, WADIFF also faces the \textbf{Challenge \#1} we proposed previously.

\noindent
\textbf{Our approach}.
To address these two challenges, we come up with some key ideas.
Specifically, to generate syntactic-correct and semantic-rich Wasm binaries, we extract AST nodes from existing real-world Wasm binaries and randomly assemble them in a syntactic-correct way. Moreover, to increase the diversity of semantics, we also import some mutations on AST-level and module-level, like introducing SIMD instructions.
As for the error locating issue, we implement a static instrumentation based error locating algorithm, which is runtime-agnostic to perform the function-level or even instruction-level root cause localization.

\section{Approach}

\begin{figure*}[t] 
    \centering 
    \includegraphics[width=1.8\columnwidth]{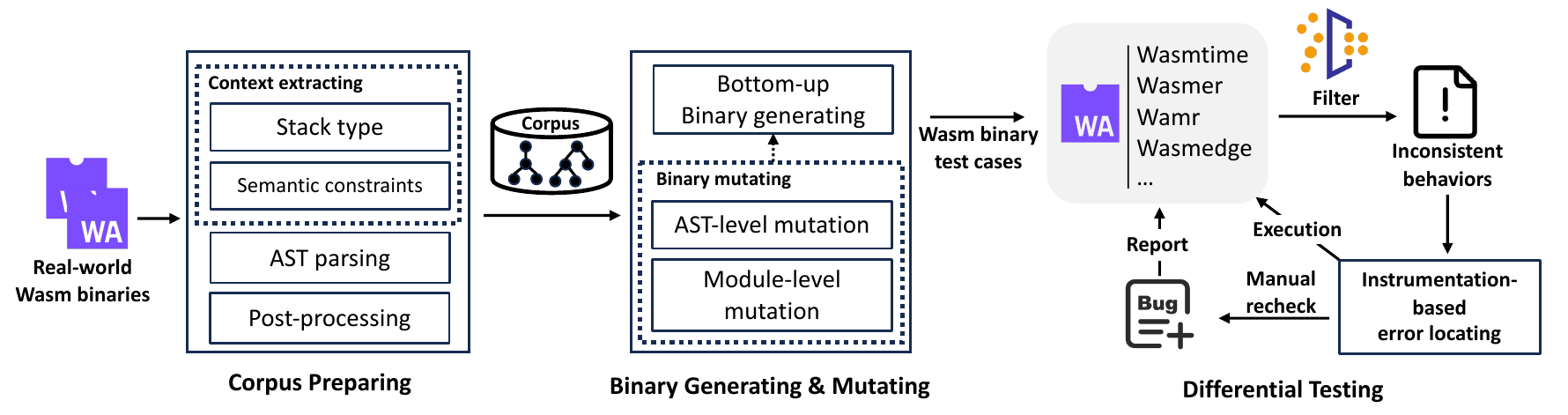} 
    \vspace{-0.2in}
    \caption{The architecture and workflow of {\framework}.} 
    \vspace{-0.1in}
    \label{fig:workflow} 
\end{figure*}

\subsection{Overview}
The workflow of {\framework} is depicted in Fig~\ref{fig:workflow}, 
which can be divided into three phases, i.e., \textit{corpus preparing}, \textit{binary generating \& mutating}, and \textit{differential testing}.
Specifically, in the corpus preparing phase, based on all collected real-world Wasm binaries, {\framework} firstly parses the corresponding ASTs and extracts valid sub-trees from them.
Then, in the binary generating \& mutating phase, {\framework} randomly assembles AST sub-trees as a valid Wasm binary. To enhance the diversity of the generated Wasm binaries, {\framework} performs AST-level and module-level mutations.
Last, in the final differential testing phase, {\framework} sends Wasm binaries to Wasm runtimes simultaneously to investigate \textit{inconsistent behaviors}. Taking advantage of static instrumentation, {\framework} can conduct a runtime-agnostic error locating on Wasm binaries that lead to such inconsistent behaviors.
We detail these three phases in the following.

\subsection{Corpus Preparing}
Instead of directly generating Wasm binaries, we decide to take real-world Wasm binaries that possess rich semantics as basic elements to assemble Wasm binaries. 
To be specific, we extract their abstract syntax tree (AST), split them into sub-trees, and assemble them in a syntactic-valid way.
Thus, the corpus preparing phase can be divided into steps including \textit{context extracting}, \textit{AST parsing}, and \textit{post-processing}, which are depicted in Algorithm~\ref{algorithm:AST}.

\renewcommand{\algorithmicrequire}{\textbf{Input:}}  
\renewcommand{\algorithmicensure}{\textbf{Output}} 

\begin{algorithm}[t]  
    \caption{The corpus preparing algorithm.}  
    \begin{algorithmic}[1] 
        \Require $instrs$ - list of instructions, $binary$ - the Wasm binary
        \Ensure $ASTs$ - the corpus, composed of a list of AST
        \Function {CorpusPreparing}{$instrs$, $binary$}  
            \State $ASTs \gets \Call{InitializeList}{ }$  
            \For{each $instr\in instrs$}
                \State $context \gets \Call{GetContext}{instr, binary}$ \Comment{\S\ref{sec:method:corpus:context}}
                \If{$instr.opcode \in [block, loop, if]$} \Comment{\S\ref{sec:method:corpus:ast}}
                    \State $node \gets Node(instr, context)$
                    \State $node.child \gets \Call{CorpusPreparing}{instr.args, binary}$
                \Else
                    \State $node \gets Node(instr, context)$
                \EndIf 
                \State $params \gets context.type.params$
                \While{$params$}
                    \For{each $preNode\in \Call{Reversed}{ASTs}$}
                        \State $node \gets \Call{AppendChild}{node, preNode}$ 
                        \State $ASTs.pop()$
                        \State $params.pop()$
                    \EndFor
                \EndWhile
                \State $ASTs.append(node)$
            \EndFor
            \State $ASTs \gets \Call{PostProcessing}{ASTs}$ \Comment{\S\ref{sec:method:corpus:postprocessing}}
            \State \Return{$ASTs$}
        \EndFunction  
    \end{algorithmic}  
    \label{algorithm:AST}
\end{algorithm} 

\noindent
\subsubsection{Context extracting.}
\label{sec:method:corpus:context}
Wasm is a statically typed language, each Wasm binary will be comprehensively verified syntactically and semantically before being executed.
Specifically, \textit{on the syntactic side}, Wasm is a stack-based language. Therefore, each instruction will consume and append a certain number of operands from and onto the stack. For example, \texttt{i32.load} requires an \texttt{i32} operand as the address and pushes the retrieved data onto the stack.
For some instructions, the number of required arguments and return values are variable, like \texttt{call} and \texttt{block}. In these cases, guaranteeing the stack balance needs to consider the signature of the callee and all included instructions, respectively.
\textit{On the semantic side}, a Wasm binary is composed of 13 sections with different functionalities.
Without considering the semantic validity, even a stack-balanced Wasm binary cannot pass the verification yet.
For example, if we have assembled a function \texttt{foo}, who invokes \texttt{bar}, we should guarantee the existence of \texttt{bar}: not only its implementation in the \textit{code section}, but also its function signature declared in the \textit{type section} and the mapping relation kept in the \textit{function section}.

\begin{table}[t]
	\caption{Instruction types and the representations, as well as the stack type and semantic constraint should be guaranteed.}
     \vspace{-0.1in}
	\resizebox{\columnwidth}{!}{%
		\begin{tabular}{lll}
			\toprule
			\textbf{Instruction}        & \textbf{Stack Type}                                      & \textbf{Semantic Constraint}                              \\ \midrule \midrule
			\multicolumn{3}{c}{Numeric Instructions}                                                                                                           \\ \midrule \midrule
			i32.const \textit{c}        & $[] \rightarrow [i32]$                                   &                                                           \\ \specialrule{0em}{0.5pt}{0.5pt} \hline \specialrule{0em}{0.5pt}{0.5pt}
			i64.add                     & $[i64, i64] \rightarrow [i64]$                           &                                                           \\
			\midrule \midrule
			\multicolumn{3}{c}{Vector Instructions}                                                                                                            \\  \midrule \midrule
			v128.const \textit{c}       & $[] \rightarrow [v128]$                                  &                                                           \\  \specialrule{0em}{0.5pt}{0.5pt} \hline \specialrule{0em}{0.5pt}{0.5pt}
			i32x4.add                   & $[v128, v128] \rightarrow [v128]$                        &                                                           \\  
			\midrule \midrule
			\multicolumn{3}{c}{Parametric Instructions}                                                                                                        \\
			\midrule \midrule
			select                      & $[t, t, i32] \rightarrow [t]$ &                                                           \\  \specialrule{0em}{0.5pt}{0.5pt} \hline \specialrule{0em}{0.5pt}{0.5pt}
			drop                        & $[t] \rightarrow []$                            &                                                           \\
			\midrule \midrule
			\multicolumn{3}{c}{Variable Instructions}                                                                                                          \\
			\midrule \midrule
			local.get \textit{n}        & $[] \rightarrow [t]$                            & \makecell[l]{$\texttt{Local} \  v$                                 \\\hspace{9pt}$\texttt{idx}_{v} = n$ \\\hspace{9pt}$\texttt{type}_{v} = t$ }                                                                                    \\ \specialrule{0em}{1pt}{1pt} \hline \specialrule{0em}{1pt}{1pt}
			global.set \textit{n}       & $[t] \rightarrow []$                            & \makecell[l]{$\texttt{Global} \  v$                                \\\hspace{9pt}$\texttt{idx}_{v} = n$ \\\hspace{9pt}$\texttt{type}_{v} = t$ }                                                                                                  \\
			\midrule \midrule
			\multicolumn{3}{c}{Memory Instructions}                                                                                                            \\ \midrule \midrule
			i32.load \textit{memarg}    & $[i32] \rightarrow [i32]$                                & \makecell[l]{$\texttt{page}_{addr} \in [min, max]$}                 \\
			\midrule \midrule
			\multicolumn{3}{c}{Table Instructions}                                                                                                             \\ \midrule \midrule
			table.get \textit{n}        & $[i32] \rightarrow [t]$                        & \makecell[l]{$\texttt{Table} \  tb$                                \\\hspace{9pt}$\texttt{idx}_{tb} = n$}  \\
			\midrule \midrule
			\multicolumn{3}{c}{Control Instructions}                                                                                                           \\  \midrule \midrule
			call \textit{n}             & $[t^*] \rightarrow [t^*]$                & \makecell[l]{$\texttt{Function} \  f$      \\$\texttt{Signature} \  s = [t^*] \rightarrow [t^*]$                        \\\hspace{9pt}$\texttt{signature}_{f} = s$  \\\hspace{9pt}$\texttt{idx}_{f} = n$ }                                                                                                 \\ \specialrule{0em}{1pt}{1pt} \hline \specialrule{0em}{1pt}{1pt}
			call\_indirect \textit{n}   & $[t^*, i32] \rightarrow [t^*]$           & \makecell[l]{$\texttt{IndirectFunction} \ f$                       \\   
 $\texttt{Signature} \  s = [t^*] \rightarrow [t^*]$   \\\hspace{9pt}$\texttt{signature}_{f} = s$  \\\hspace{9pt}$\texttt{idx}_{s} = n$  }                                                                                                \\ \specialrule{0em}{1pt}{1pt} \hline \specialrule{0em}{1pt}{1pt}
			block \textit{n}            & $ [t^*] \rightarrow [t^*]$                 & \makecell[l]{ $\texttt{Signature} \  s = [t^*] \rightarrow [t^*] $ \\\hspace{9pt}$\texttt{idx}_{s} = n$}                                                                                                  \\
			\bottomrule
		\end{tabular}%
	}
     \vspace{-0.2in}
	\label{table:context}
\end{table}

To solve this challenge, we firstly categorize Wasm instructions into seven groups according to their operations on the stack, i.e., \textit{stack type}, and the corresponding must-satisfied \textit{semantic constraints}. Table~\ref{table:context} illustrates some representatives of each group.
Specifically, the stack type of numeric instructions is fixed, and there are no extra semantic constraints on them.
Vector instructions are similar to them, but are specifically designed for the vector type (see \S\ref{sec:backgroud:wasm}).
In addition, the stack type of parametric instructions may vary. For example, \texttt{drop} consumes an operand from the stack whichever its type. Thus, the stack type of \texttt{drop} is $[t] \rightarrow []$, where \textit{t} is a wildcard and refers to \textit{any type}.
The other four types of instructions need semantic constraint.
For example, $\texttt{local.get} \ n$ has two pieces of constraints: for a local variable $v$, its index and type should be $n$ and $t$, respectively.
As for memory and table instructions, they are respectively responsible for interacting with the memory and table area. They also limited by the corresponding semantic constraints. For instance, \texttt{i32.load} requires the target address must be within the valid address range, i.e., $[min, max]$. Table instructions require the corresponding index should be existed and matched.
In Wasm, two kinds of function invocations exist, through \texttt{call} and \texttt{call\_indirect}, respectively. They have subtle differences in semantic constraints. For \texttt{call} \textit{n}, it directly invokes the \textit{n}-th function,
while the \textit{n} in \code{call\_indirect} \textit{n} indicates the index of the function type declared in the type section. The index of callee is retrieved from the stack, i.e., the second \texttt{i32} of the parameters of its stack type.

To build the corpus, the algorithm takes a list of instructions and its corresponding binary as input. While going through the instruction list (L3), the algorithm extracts context information at L4.
Specifically, for each instruction, it not only concretizes the stack type according to the context provided by the binary, but also binds the corresponding semantic constraints to the instruction.
For example, when encountering a \texttt{call} instruction, whose number and type of consumed elements on the stack are not statically determined, the algorithm extracts its immediate number $n$ and obtains its function signature by indexing through the function and type section. To this end, the $[t^*] \rightarrow [t^*]$ can be concretized, like $[i32, i32] \rightarrow [i64]$.
In summary, the algorithm can bind the necessary context information to each instruction according to stack type and semantic constraints declared in Table~\ref{table:context}.

\subsubsection{AST parsing.}
\label{sec:method:corpus:ast}
Parsing ASTs of real-world Wasm binaries is necessary before extracting sub-trees from them to build the corpus.
Taking advantage of the extracted context as we mentioned in \S\ref{sec:method:corpus:context}, a Wasm binary can be easily parsed to the corresponding AST.
To better illustrate how this process happens, Fig.~\ref{fig:subfiga} to Fig.~\ref{fig:subfigc} show a factorial function written in C, the instruction list of the compiled Wasm binary, and the corresponding AST, respectively. We detail the process combining with Algorithm~\ref{algorithm:AST}.
As we can see from Fig.~\ref{fig:subfigb}, the compiled instruction list is quite flat, which is hard to identify the AST structure.
Therefore, the algorithm firstly identifies if current instruction is any of \texttt{block}, \texttt{loop}, or \texttt{if} (L5 to L7 in Algorithm~\ref{algorithm:AST}) to recursively build AST as only sub-trees led by these instructions can be nested.
Then, the algorithm constructs the AST according to the stack type, as shown from L11 to L20 in Algorithm~\ref{algorithm:AST}.
For example, in the instruction list, the first instruction, i.e., \texttt{i32.const 5}, takes no elements from the stack, thus it jumps over the iteration at L12 and is pushed to $ASTs$ at L19.
Then, its following \texttt{local.set 0} takes an element from the stack as shown in Table~\ref{table:context}. Thus, the while-loop at L12 is executed once. Within the while-loop, the last AST node, i.e., the previous \texttt{i32.const 5} is set as the child for \texttt{local.set 0}, which will then be pushed to $ASTs$.
Even the algorithm meets \texttt{call}, such an instruction with variable number of arguments, its context is determined in \S\ref{sec:method:corpus:context} by retrieving the callee's function signature, thus the AST can be built without any issue.
Due to the recursive construction as shown at L7 in Algorithm~\ref{algorithm:AST}, the \texttt{block} (L7 of Fig.~\ref{fig:subfigb}) will lead the \texttt{loop} (L8 of Fig.~\ref{fig:subfigb}), within which it implements the factorial calculation as shown in the C source code.
Consequently, after traversing the \texttt{factorial} function, $ASTs$ is composed of four nodes, each of which can be regarded as a root of a sub-tree of the AST (Fig.~\ref{fig:subfigc}).

\subsubsection{Post-processing}
\label{sec:method:corpus:postprocessing}
\begin{figure}[t]
	\centering
	\begin{minipage}[c]{0.55\columnwidth}
		\raggedright
		\includegraphics[width=\columnwidth]{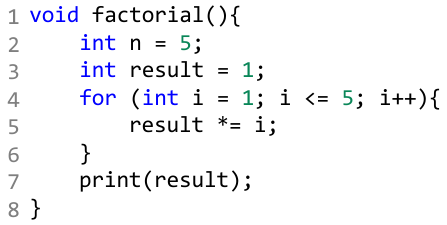}
		\subcaption{The source code in C.}
		\label{fig:subfiga}
	\end{minipage} 
	\begin{minipage}[c]{0.3\columnwidth}
		\raggedleft
		\includegraphics[width=\columnwidth]{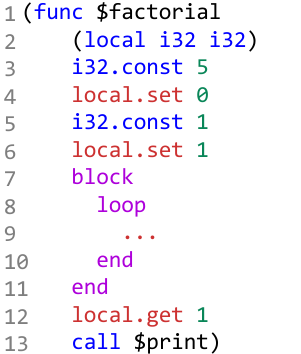}
		\subcaption{The compiled bytecode.}
		\label{fig:subfigb}
	\end{minipage} \\
	\begin{minipage}[c]{0.85\columnwidth}
		\centering
		\includegraphics[width=\columnwidth]{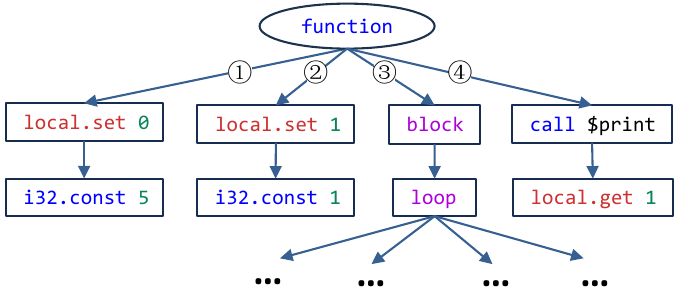}
        \vspace{-0.1in}
		\subcaption{The parsed AST of the Wasm bytecode.}
		\label{fig:subfigc}
	\end{minipage}
     \vspace{-0.1in}
	\caption{A concrete example of AST parsing.}
     \vspace{-0.2in}
	\label{fig:mainfig}
    \vspace{-0.05in}
\end{figure}

As we can see from Fig.~\ref{fig:subfigc}, the 1st and 2nd sub-trees are quite similar. In other words, if we take both of them into valid elements in the final corpus, the generated Wasm binaries may be duplicated.
Therefore, in the post-processing stage, we try to minimize the size of the corpus to avoid generating duplicated Wasm binaries.
In general, for each sub-tree, we firstly remove the immediate number of each instruction. Then, we adopt a DFS on sub-trees to identify the ones with identical instruction dependent relationship.
Take Fig.~\ref{fig:subfigc} as an example, because both 1st and 2nd sub-trees have \texttt{local.set} as the root node of \texttt{i32.const}, only one of them will be retained.

\subsection{Binary Generating \& Mutating}
\label{sec:method:binary}
Given the sub-trees extracted from ASTs of real-world Wasm binaries as corpus, as well as the extracted context information on each instruction, {\framework} can generate valid Wasm binaries efficiently and effectively. Additionally, {\framework} also conducts mutations on these generated binaries to bring in more diversities.
The \textit{binary generating} and \textit{binary mutating} are detailed as follows.

\subsubsection{Binary generating}
\label{sec:method:binary:generating}
We generate a Wasm binary following a \textit{bottom-up} way, i.e., \textit{building a valid entry function}, \textit{maintaining necessary invoking relation}, and \textit{supplementing extra semantics}.

\noindent
\textbf{Step I: Building a valid entry function.}
First of all, we need to generate a function body of an entry function.
In general, we randomly sample a specified number of AST sub-trees from the corpus. Then, we concatenate and transform them into a sequence of Wasm instructions, which is regarded as the body of the entry function.
In Wasm, a function is composed of not only a set of instructions as its implementation, but also some local variables that can be accessed within the current function.
Thus, {\framework} adds five local variables for the entry function, typed as \texttt{i32}, \texttt{i64}, \texttt{f32}, \texttt{f64}, and \texttt{v128} (see \S\ref{sec:backgroud:wasm}) and indexed from 0 to 4, respectively.
For each variable instruction in the newly generated entry function, {\framework} gets their stack type mentioned in Table~\ref{table:context} and conducts necessary rewriting on the immediate number.
For example, if a \texttt{local.get} is bound an operand with type \texttt{i64}, its immediate number will be rewritten as 1.
Finally, according to the signature of the entry function, {\framework} appends necessary \texttt{local.get} before the final \texttt{return} instruction.
Consequently, {\framework} generates an entry function composed of five local variables and a list of instructions, where the immediate number of each variable instruction is rewritten to the index according to its bound stack type extracted in Algorithm~\ref{algorithm:AST}.

\begin{figure}[t] 
    \centering 
    \includegraphics[width=0.8\columnwidth]{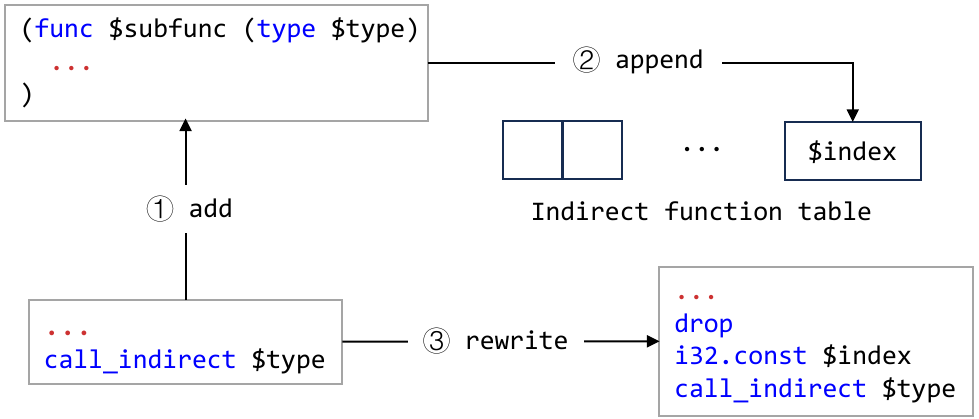} 
        \vspace{-0.1in}
    \caption{The concrete steps of {\framework} on maintaining necessary invoking relation of the \texttt{call\-indirect} instruction.} 
    \vspace{-0.1in}
    \label{fig:indirect} 
\end{figure}

\noindent
\textbf{Step II: Maintaining necessary invoking relation.}
Except for considering the validity intra-functionally when building the entry function, the inter-functional validity should also be ensured, i.e., maintaining function invoke relations from the entry function.
In Wasm, only two instructions can invoke function calls, i.e., \texttt{call} and \texttt{call\_indirect}.
Thus, {\framework} traverses each instruction. When meets \texttt{call}, {\framework} generates a callee according to its bound concretized stack type and semantic constraint.
Instead of generating a callee with an empty function body whose aim is solely maintaining the stack balance, the callee is adopted the same method as we mentioned in the \textbf{Step I}. In other words, the callee has a functional function body, and it may recursively generate its callees.
As for \texttt{call\_indirect}, maintaining its invoking relation requires a little extra effort. Specifically, the callee of \texttt{call\_indirect} is determined at runtime.
As shown in Fig~\ref{fig:indirect}, {\framework} firstly generates a callee as its handling on \texttt{call}.
Then, it inserts the function index into the function table to make the indexing process raise no exceptions.
Last, to make sure the \texttt{call\_indirect} can actually be guided to the newly generated callee, {\framework} inserts a \texttt{drop} and an \texttt{i32.const} \textit{c} before it, where the \textit{c} is the index.

\noindent
\textbf{Step III: Supplementing extra semantics.}
For a Wasm binary, only focusing on functions cannot fully guarantee its semantic correctness because the functionalities are decoupled into all 13 sections.
Thus, extra semantics should be supplemented.
Take the \textit{data} section as an example, which is used to initiate the linear memory of Wasm binary. To avoid memory-related instructions from raising exceptions, {\framework} goes through the memory instructions of each function to determine the maximum addressing range depending on its semantic constraint. Within the range, {\framework} will fill random data into the \textit{data section}.
Similarly, for the \textit{table} and \textit{element} sections that make up the indirect function table of Wasm, we construct them based on the context of table instructions in the binary.
This ensures correct executions on table instructions.
As for the global section, whose elements can be accessed by all functions, we adopt the same strategy as inserting local variables in functions. That is, {\framework} only inserts five global elements with five different type in the global section, and rewrites the immediate numbers of all \texttt{global.get} and \texttt{global.set} instructions according to their bound stack type to the corresponding index.

\subsubsection{Binary mutating}
\label{sec:method:binary:mutating}

Though a substantial number of Wasm binaries can be generated through the methods we mentioned in \S\ref{sec:method:binary:generating}, there are two shortcomings.
On the one hand, it is still difficult to increase the code coverage of tested runtimes through assembling ASTs from existing Wasm binaries.
On the other hand, due to the ongoing evolution of Wasm, it is hard to verify the implementation of runtimes on new features. For instance, SIMD instructions is newly introduced to handle vectors, and the number of SIMD related instructions is 236, far more than instructions defined in Wasm 1.0~\cite{wasm1}.

To address these shortcomings, we propose a set of mutation strategies to try to comprehensively test Wasm runtimes and uncover hidden issues in them.
Generally speaking, the mutation strategies can be divided into \textit{AST-level mutation} and \textit{module-level mutation}, while the former one can be integrated into the \textbf{Step I} and \textbf{Step II} in \S\ref{sec:method:binary:generating} and the latter one can be performed when conducting the \textbf{Step III} in \S\ref{sec:method:binary:generating}.

\noindent
\textbf{AST-level mutation} 
is to mutate the instructions of ASTs in the corpus we collected, and its purpose is to extend the semantics of AST, so as to further improve the code coverage of runtime testing.

\begin{itemize}[leftmargin=*]
\item \textit{Mutate immediate numbers.} Improper processes on corner cases are more likely to trigger bugs or even vulnerabilities~\cite{zhang, wave, mswasm}.
Therefore, part of the mutation resources is put on the immediate number of instructions.
For example, for constant instructions, like \texttt{i32.const}, {\framework} tries to replace the original operand with the value near the boundary, e.g., $2^{32}-1$.
Furthermore, for memory load and store instructions, {\framework} attempts to mutate the offset and alignment arguments.

\item \textit{Mutate operations.} The remaining mutation resources are put on the operations themselves of instructions.
Generally speaking, to support SIMD instructions, {\framework} conducts mutations on numeric and memory instructions to the SIMD ones with similar semantics.
For example, in Fig~\ref{fig:convert}, the original implementation (left side) is \texttt{i32.const} followed by an \texttt{i32.load}. After mutating both of them into SIMD instructions with similar semantics, the mutated implementation (right side) is \texttt{v128.const} followed by a \texttt{v128.load}. However, we can observe that the type of the consumed element of \texttt{v128.load} mismatches the one pushed by \texttt{v128.const}. Fortunately, Wasm specification provides a set of instructions to convert the element type~\cite{instrs}. Thus, {\framework} adds an extra \texttt{i32x4.splat} to convert \texttt{v128} to \texttt{i32} to make the mutated implementation valid.

We further design a strategy to increase the diversity of ASTs.
Specifically, as illustrated in Table~\ref{table:context}, instructions may have identical stack type, like \texttt{i32.add} and \texttt{i32.sub}. Arbitrarily conducting interchanges among these instructions will not break the stack balance. Therefore, {\framework} randomly replaces instructions to others according to the stack type.

\begin{figure}[t] 
    \centering 
    \includegraphics[width=0.8\columnwidth]{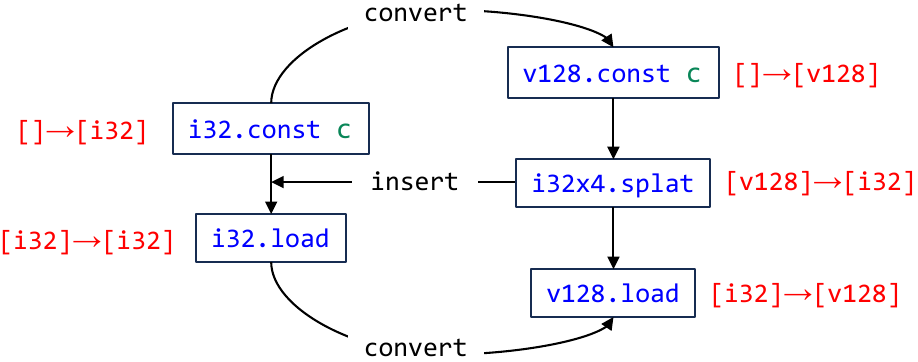} 
        \vspace{-0.1in}
    \caption{Mutating normal instructions to the SIMD ones.} 
        \vspace{-0.1in}
    \label{fig:convert} 
\end{figure}

\end{itemize}

\noindent
\textbf{Module-level mutation.} 
In addition to the inconsistency in the implementation of instructions, deficiencies in the runtime's verification on the whole Wasm module can also pose security risks. For example, incomplete memory boundary checks may lead to sandbox escapes.
Thus, we also perform module-level mutations during the \textbf{Step III} in \S\ref{sec:method:binary:generating}.
Specifically, Cao et al.~\cite{brewasm} has summarized five functionalities in Wasm, i.e., \textit{Global}, \textit{Import \& Export}, \textit{Memory}, \textit{Function}, and \textit{Custom}.
Except for the last functionality, which has no association with the semantics of a Wasm binary, against each of the others, we have designed a specific mutation strategies, which are detailed in the following.

\begin{itemize}[leftmargin=*]
	\item \textit{Global.} Except for the data type, each global variable possesses various other attributes, such as whether the variable is mutable. We randomly mutate the attributes of each global element.
	\item \textit{Import \& Export.} 
 Wasm can import or export various semantics, such as functions, memory, and even global variables. {\framework} randomly adds import and export items (in the import and export sections) to the generated binaries. 
	\item \textit{Memory.} As we mentioned in the \textbf{Step III} in \S\ref{sec:method:binary:generating}, the data section is initiated with random data. Except for that, Wasm still requires all memory accesses should be within the valid address limitation. Thus, {\framework} mutates the address limitation to test the correctness of runtime's memory boundary checks.
	\item \textit{Function.} Part of this functionality is related to the code section, which is handled by \S\ref{sec:method:binary:generating}. Thus, {\framework} primarily mutates the table section. Similar to the memory section, the table section determines the range of the indirect function table, which is used for indexing by \texttt{call\_indirect}. To this end, the correctness of runtime's boundary checks can be covered.
\end{itemize}

\subsection{Differential Testing}
\label{sec:approach:testing}
During differential testing, we firstly should identify the inconsistency among runtimes. Additionally, due to the large number of Wasm binaries being fed to runtimes, many inconsistencies might be caused by the same underlying issue. Effectively locating the root cause can improve the efficiency of the whole testing process. We next explain how we address these issues.

\subsubsection{Inconsistency Identification}
\label{sec:approach:testing:inconsistency}
During the differential testing process, we \textit{cannot simply take different output as inconsistency} because some of them are triggered by non-bug factors, like differences in design principles and implementational distinctions. 
First, Wasm is still under development, thus runtimes may have different levels on supporting the Wasm specification.
For example, some runtimes may not support SIMD instructions yet, resulting in an unsupported prompt when encountering Wasm binaries with SIMD instructions. 
Second, the output of various runtimes is also influenced by their implementation styles. For example, some runtimes may consider \texttt{i32} and \texttt{i64} integers as signed when outputting them, while other runtimes may treat them as unsigned. The Wasm specification does not explicitly define these behaviors.
Third, different runtimes adopt diverse ways to handle error messages. For example, when encountering an out-of-bounds table access, some runtimes may only output \textit{undefined element}, while other runtimes provide more specific reasons, such as \textit{out of bounds table access}. 

\begin{figure}[t] 
    \centering 
    \includegraphics[width=0.6\columnwidth]{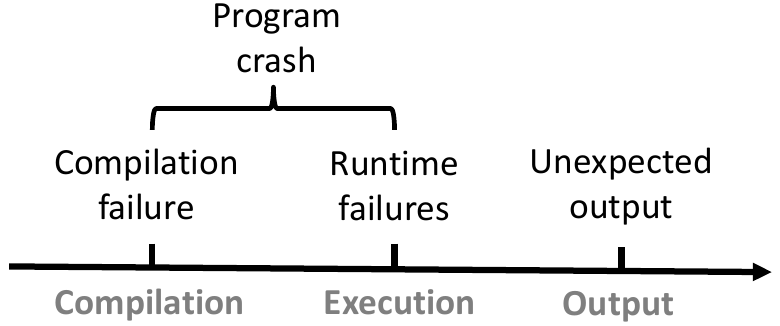} 
        \vspace{-0.1in}
    \caption{The types of inconsistencies during execution.} 
        \vspace{-0.1in}
    \label{fig:classification} 
\end{figure}

To reduce the number of false positives, i.e., inconsistent behaviors due to the above non-bug reasons, we give a clear definition of \textit{inconsistent behavior} as shown in Fig.~\ref{fig:classification}.
Specifically, we define three kinds of inconsistent behaviors based on the lifetime of a Wasm binary being executed by runtimes, namely \textit{compilation failure} (\textbf{CF}), \textit{runtime failure} (\textbf{RF}), and \textit{unexpected output} (\textbf{UO}). These three types are mutually exclusive and cover all inconsistent behaviors we identified during the differential testing.
The first two types happen when runtimes encounter unexpected errors during the corresponding stage.
As the name suggests, the compilation failure occurs when the Wasm binary is compiled and instantiated. The Wasm runtime will crash, that is, the input Wasm binary will not be executed at all.
When Wasm binaries are executed, runtimes may still crash, usually due to raised exceptions, which is named as runtime failure. We identify this inconsistency behavior by the classes of raised exceptions.
The last one, i.e., unexpected output, can only be observed when a Wasm binary is executed.

As for identifying which runtime indeed has an unexpected behavior, we take the majority rule as other differential testing studies adopt~\cite{wadiff}, i.e., we consider the behavior generated by most runtimes when executing a binary as correct, while runtimes exhibiting other types of behavior are considered potentially buggy.
For example, if a binary encounters out-of-bounds memory access in three runtimes and another runtime prompts an indirect function table related error, the latter runtime is considered potentially buggy.

\subsubsection{Root Cause Locating}
\label{sec:approach:testing:locating}
Different Wasm binaries may trigger the same bug in Wasm runtimes, thus performing an error locating is necessary to deduplicate these Wasm binaries according to the root causes of triggered inconsistency. 
To this end, we propose a \textit{runtime-agnostic binary instrumentation method for root cause localization}, which is detailed in Algorithm~\ref{algorithm:locating}, consisting of \textit{function-level localization} and \textit{instruction-level localization}.

\begin{algorithm}[t]
    \caption{The error locating algorithm.}  
    \label{algorithm:locating} 
    \begin{algorithmic}[1] 
        \Require $binary$ - inconsistent binary, $type$ - inconsistent type
        \Ensure $funcid$ - the index of inconsistent function. $instr$ - the inconsistent instruction
        \Function {FuncLocating}{$binary, type$}
            \State $logsList \gets \Call{FuncInstrumentation}{binary}$
            \State $funcid \gets None$
            \For{each $logs\in logsList$}
                \If{$\Call{CompareFuncLogs}{logs} = False$}
                    \State $funcid \gets \Call{FindInconsistentFunc}{logsList}$
                \EndIf
            \EndFor
            \If{$type = OUTPUT$}
                \State $instr \gets InstrLocating(binary, funcid)$
                \State \Return{$instr$}
            \Else
                \State \Return{$funcid$}
            \EndIf
        \EndFunction
    \end{algorithmic}
    \begin{algorithmic}[1] 
        \Require $binary$ - inconsistent binary, $funcid$ - the index of inconsistent function
        \Ensure $instr$ - The inconsistent instruction
        \Function {InstrLocating}{$binary, funcid$}
            \State $logsList \gets \Call{InstrInstrumentation}{binary, funcid}$
            \State $instr \gets None$
            \For{each $logs\in logsList$}
                \If{$\Call{CompareInstrLogs}{logs} = False$}
                    \State $instr \gets logs.instr$
                    \State $break$
                \EndIf
            \EndFor
            \State \Return{$instr$}
        \EndFunction
    \end{algorithmic}
\end{algorithm}

\noindent
\textbf{Function-level Localization.}
It is designed for \textit{runtime failure} and \textit{unexpected output} inconsistency types (see \S\ref{sec:approach:testing:inconsistency}).
Specifically, the localization is implemented by static binary instrumentation. Before and after each \texttt{call} instruction, the instrumented function is responsible for printing the index of the invoked function along with its parameters or return values, which is achieved by the imported \code{fd\_write} function, one of the WASI functions (see \S\ref{sec:backgroud:runtime}).
Formally, the \texttt{FuncLocating} in Algorithm~\ref{algorithm:locating} illustrates the implementation of function-level localization.
It takes the binary that exhibits inconsistencies and its corresponding inconsistency type as inputs. Then the binary is instrumented at the function level and then executed by runtimes to obtain the respective output logs (L2).
At the loop at L4, the algorithm iterates over the logs outputted by instrumented functions, consisting of arguments and return values. 
The algorithm then compares the logs among runtimes to find the function that caused the inconsistent behavior (L5-L7). It is worth noting that the binary generating method guarantees the call relationship is quite simple, i.e., each function is only called once and there is not recursive invocation, thus we can find the inconsistent function with a simple queue traversing (L6).
Lastly, the algorithm examines whether the instruction-level localization should be further invoked depending on if the inconsistency type is unexpected output.

\noindent
\textbf{Instruction-level Localization.}
The \texttt{InstrLocating} will be invoked if the inconsistency behavior is due to the unexpected output.
Specifically, it takes the binary exhibiting inconsistencies and the index of its inconsistent function as input. Then \texttt{InstrLocating} performs instruction-level instrumentation on the specified function within the binary and feeds the instrumented binary to runtimes to get their output logs (L2).
The instrumentation is conducted by printing the opcode of non-control-flow instruction and the value on the top of the stack after each instruction.
L4 to L10 iterate the output of each instrumentation point. By comparing the values on the stack after the execution of each instruction, the algorithm identifies the specific instruction responsible for the observed inconsistency behavior.
Consequently, the instruction that leads to buggy inconsistency is returned.

\section{Implementation \& Evaluation}

\subsection{Implementation \& Experimental Setup}
\label{sec:imple-eval:imple}
We have implemented {\framework} with over 7.4K LOC of Python3 code from scratch. All experiments were performed on a server in Ubuntu 22.04 with a 64-core AMD EPYC 7713 CPU and 256GB RAM. Our evaluation is driven by the following research questions:

\begin{itemize}
    \item[\textbf{RQ1}] What is the performance of {\framework} compared with baselines?
    \item[\textbf{RQ2}] How many real-world bugs in Wasm runtimes can be identified by {\framework}?
    \item[\textbf{RQ3}] What are the characteristics of the detected bugs?
\end{itemize}

\noindent
\textbf{Benchmark.}
We use the WasmBench~\cite{wasmbench_github}, a well-known benchmark consisting of over 8K Wasm binaries. These binaries are collected from various sources, including code repositories, web applications, and package managers. 
Consequently, it guarantees the richness in terms of semantics of Wasm binaries, laying the foundation for {\framework} to generate a corpus.

\noindent
\textbf{Baselines.}
\label{sec:evaluation:baseline}
We select wasm-smith~\cite{wasm-smith} and WADIFF~\cite{wadiff} as baselines.
Specifically, wasm-smith is a Wasm binary generator that is widely-adopted in testing Wasm runtimes~\cite{wasmtime,wasmer}. It is worth noting that Wasm binaries generated by wasm-smith in the default mode are mostly unable to be directly executed by runtimes because it may import functions that are supported by runtimes. Therefore, we configure it to generate Wasm binaries without imported functions, and each binary has a minimum of ten functions, all of which are exported. 
As for WADIFF, it adopts symbolic execution on specifications of each instruction to generate Wasm binaries to test the correctness of Wasm runtimes on executing instructions.

\noindent
\textbf{Targeted Runtime.}
We select the representative Wasm runtimes according to two criteria. First, the stars of the runtime are greater than 3K. Second, the runtime is actively maintained for the last three months and has been officially released for over a year. 
Consequently, wasmtime~\cite{wasmtime}, Wasmer~\cite{wasmer}, wamr~\cite{wamr}, and WasmEdge~\cite{wasmedge} are selected. Note that, no inconsistent behaviors are identified for Wasmer, thus we omit it in the following three RQs.

\begin{table}[]
\centering
\caption{The number of generated Wasm binaries and the ones that can trigger inconsistent behaviors, where \textbf{CF}, \textbf{RF}, \textbf{UO}, and \textbf{T} refer to compilation failure, runtime failure, unexpected output, and the sum of these cases, respectively. \textbf{B} and \textbf{UB} represent to bugs and unique bugs that are not discovered by the other two tools.}
    \vspace{-0.1in}
\label{table:rq1}
\resizebox{0.8\columnwidth}{!}{%
\begin{tabular}{@{}ccccc@{}}
\toprule
\multicolumn{2}{c}{}                                                                              & \textbf{{\framework}} & \textbf{wasm-smith}  & \textbf{WADIFF}      \\ \midrule
\multicolumn{2}{c}{\textbf{\begin{tabular}[c]{@{}c@{}}\# Generated\\ Wasm Binaries\end{tabular}}} & 269,998               & 343,299              & 283,330              \\ \midrule
\multirow{6}{*}{\textbf{wasmtime}}                          & \textbf{CF}                         & 3,151                 & 0                    & 0                    \\
                                                            & \textbf{RF}                         & 0                     & 0                    & 0                    \\
                                                            & \textbf{UO}                         & 0                     & 0                    & 0                    \\
                                                            & \textbf{T}                          & 3,151                 & 0                    & 0                    \\
                                                            & \textbf{B}                          & 1                     & 0                    & 0                    \\
                                                            & \textbf{UB}                         & 1                     & 0                    & 0                    \\ \midrule
\multirow{6}{*}{\textbf{wamr}}                              & \textbf{CF}                         & 40,158                     & 5,555                    &   0                  \\
                                                            & \textbf{RF}                         & 7,578                      & 5,479                    & 0                     \\
                                                            & \textbf{UO}                         & 1,441                      &  0                & 443                  \\
                                                            & \textbf{T}                          & 49,177                & 11,034               & 443                  \\
                                                            & \textbf{B}                          & 7                     & 5                    & 2                    \\
                                                            & \textbf{UB}                         & 4                     & 4                    & 0                    \\ \midrule
\multirow{6}{*}{\textbf{WasmEdge}}                          & \textbf{CF}                         & 0                     & 0                    & 0                    \\
                                                            & \textbf{RF}                         & 10,032                     & 0                    & 0                    \\
                                                            & \textbf{UO}                         & 3,543                      & 0                    & 0                    \\
                                                            & \textbf{T}                          & 13,575                & 0                    & 0                    \\
                                                            & \textbf{B}                          & 7                     & 0                    & 0                    \\
                                                            & \textbf{UB}                         & 7                     & 0                    & 0                    \\ \midrule
\multirow{6}{*}{\textbf{\begin{tabular}[c]{@{}c@{}}Total\\ (Sum of\\ above\\ runtimes)\end{tabular}}} & \textbf{CF} & 43,309  & 5,555 & 0 \\
                                                            & \textbf{RF}                         & 17,610  & 5,479 & 0 \\
                                                            & \textbf{UO}                         & 4,984  & 0 & 443 \\
                                                            & \textbf{T}                          & 65,903 & 11,034 & 443 \\
                                                            & \textbf{B}                          & 15  & 5 & 2 \\
                                                            & \textbf{UB}                         & 12  & 4 & 0 \\ \bottomrule
\end{tabular}%
}
    \vspace{-0.2in}
\end{table}

\subsection{RQ1: Comparison with Baselines}
\label{sec:evaluation:rq1}

\noindent \textbf{Overall Result.}
For fair comparison,
we run each tool separately for 24 hours on all four targeted Wasm runtimes.
Table~\ref{table:rq1} illustrates the number of generated Wasm binaries (the second row) and the ones that are able to result in inconsistent behaviors (the rows leading by \textbf{T}).
As we can see, the efficiency of these three tools is indistinguishable, and they all can generate and perform the corresponding differential testing on 300,000 Wasm binaries within 24 hours.
However, when it comes to the number of Wasm binaries that can lead to inconsistent behaviors, the difference in efficiency among them is significant. {\framework} generate more than 65K Wasm binaries that can result in inconsistent behaviors, which is 6.0x and 148.8x than wasm-smith and WADIFF respectively.
Interestingly, we can easily observe that wasm-smith and WADIFF can only detect inconsistent behaviors in wamr, while {\framework} is also effective in wasmtime and WasmEdge.
This is because the aim of binaries generated by wasm-smith is more about testing the syntax verification of runtimes, which may has been widely tested by the other runtimes before release.
As for WADIFF, it can only test the implementation inconsistency of instruction-level.

\noindent \textbf{Distribution of inconsistencies.}
We further analyze the types of inconsistencies (mentioned in \S\ref{sec:approach:testing:inconsistency}) identified,
as the rows led by \textbf{CF} (compilation failure), \textbf{RF} (runtime failure), and \textbf{UO} (unexpected output) in Table~\ref{table:rq1}.
Interestingly, for wasmtime, all identified inconsistency behaviors are \textbf{CF}, while all cases of WasmEdge fall into another two categories.
We speculate that this is due to the different robustness of the two runtimes in their modular implementations.
Moreover, the most cases that can lead to inconsistent behavior can be found in wamr for all the three tools, while only {\framework} can identify all three kinds of inconsistencies, which further suggests the effectiveness of {\framework}.

\noindent \textbf{Bugs.}
We make a step further to investigate the 
root causes of these inconsistency behaviors, as shown in the rows led by \textbf{B} and \textbf{UB}, indicating the number of bugs and unique bugs that are not discovered by the other two tools. 
{\framework} identifies 15 bugs in total within 24 hours, while 12 of which are not discovered by wasm-smith and WADIFF. Interestingly, we can also observe that wasm-smith found four unique bugs that are not identified by {\framework} during testing against wamr, which are all syntactic bugs.
For example, one case has a \texttt{br\_table} followed by an array with more than 30 branches as targets, where wamr failed to examine if the type of each targeted block is identical.
Such a long \texttt{br\_table} is unusual in real-world cases, thus {\framework} cannot find this issue within 24 hours even with the help of mutation strategies.

\textbf{Answer to RQ1.} \textit{Compared with the state-of-the-art baselines, i.e., wasm-smith and WADIFF, {\framework} illustrates its superiority in both terms of efficiency and effectiveness in identifying unexpected behaviors in Wasm runtimes, where the number of identified cases is 6.0x and 148.8x than them, respectively.}

\subsection{RQ2: Runtime Bugs in the Wild}
\label{sec:evaluation:rq2}
Our previous exploration suggests that {\framework} outperforms SOTA techniques greatly. Thus, we next seek to apply {\framework} to identifying runtime bugs in the wild.
For the targeted runtimes, {\framework} performs differential testing for 72 hours\footnote{Note that, the reason we set the time frame of 72 hours is that the number of unique bugs {\framework} identified remain stable after roughly 65 hours.}.
In total, {\framework} has generated 832,053 Wasm binaries, 167,343 out of which (accounting for 20.11\%) can lead to inconsistent behavior across Wasm runtimes.
Table~\ref{table:rq2} illustrates the break down details.

\begin{table}[t]
\caption{The break down of 167,343 Wasm binaries that can lead to inconsistent behaviors generated by {\framework}. The number in parenthesises is the number of unique bugs.}
    \vspace{-0.1in}
\label{table:rq2}
\resizebox{0.9\columnwidth}{!}{%
\begin{tabular}{rcccc}
\toprule
                    & \makecell{\textbf{CF}}  & \makecell{\textbf{RF}} & \makecell{\textbf{UO}} & \makecell{\textbf{Total}} \\
\midrule
wasmtime       & 8,692 (1)    & 0 (0)         & 0 (0)  &  8,692 (1)\\
wamr      & 99,578 (4)    & 17,568 (10)       & 2,763 (5)  &  119,909 (19)  \\
WasmEdge & 0 (0)    & 30,273 (7)        & 8,469 (6)    &  38,742 (13) \\
\textbf{Total} & 108,270 (5) & 47,841 (17) & 11,232 (11) & 167,343 (33) \\
\bottomrule
\end{tabular}%
}
    \vspace{-0.1in}
\end{table}

\noindent \textbf{Behavior Inconsistencies.}
Among all these three runtimes, {\framework} generates a total of 119,908 cases that can trigger inconsistencies in wamr, which is 3.1x and 13.8x than WasmEdge and wasmtime. Most of the inconsistent cases belong to \textbf{CF}, accounting for 65\% of all binaries.

\noindent \textbf{Bugs.}
As shown in Table~\ref{table:rq2}, for all the 167,343 inconsistent cases, our root cause localization method pinpoints 33 unique bugs, and 5, 17, and 11 of them are classified into \textbf{CF}, \textbf{RF}, and \textbf{UO}, respectively.
Over half of the bugs are discovered within the first 24 hours, and the growth rate gradually slows down in the following hours.
Note that, against wasmtime, we only identified a single type bug that can generate its unexpected behaviors, which is consistent with the results presented in RQ1.
It is noteworthy that although {\framework} generated the highest number of binaries resulting in \textbf{CF}, there are only 4 unique bugs.
After investigating the reason, we found that it is because wamr raises exceptions when parsing the complex representation of immediate values of \texttt{v128.const}. Because our mutation strategies (see \S\ref{sec:method:binary:mutating}) can import such instructions, a large number of binaries trigger this inconsistent behavior.

\begin{table}[t]
	\centering
	\caption{Detailed information of all confirmed bugs.}
     \vspace{-0.1in}
	\label{table:bugs}
	\resizebox{\columnwidth}{!}{%
		\begin{tabular}{llccc}
			\hline
			\textbf{Runtime}                & \textbf{Issue}  & \textbf{Type}           & \textbf{Root Cause}                       & \textbf{Status}    \\ \hline
			wasmtime               & \#7558 & \textbf{CF} & type conversion error            & Fixed     \\ \hline
			\multirow{11}{*}{wamr} & \#2450 & \textbf{RF} & integer overflow                 & Fixed     \\
			                       & \#2555 & \textbf{RF} & llvm compiler crash               & Fixed     \\
			                       & \#2556 & \textbf{CF} & non-alignment issue               & Fixed     \\
			                       & \#2557 & \textbf{CF} & value kind check missing          & Fixed     \\
			                       & \#2561 & \textbf{UO} & instruction implementation error  & Fixed     \\
			                       & \#2677 & \textbf{CF} & type conversion error             & Fixed     \\
			                       & \#2690 & \textbf{UO} & instruction implementation error  & Fixed     \\
			                       & \#2720 & \textbf{RF} & memory check error                & Fixed     \\
			                       & \#2789 & \textbf{CF} & string check error                & Fixed     \\
			                       & \#2861 & \textbf{UO} & instruction implementation error & Confirmed \\
			                       & \#2862 & \textbf{UO} & instruction implementation error & Confirmed \\
			\hline
			\multirow{14}{*}{WasmEdge}
			                       & \#2812 & \textbf{UO} & instruction implementation error & Confirmed \\
			                       & \#2814 & \textbf{RF} & memory check error               & Confirmed \\
			                       & \#2815 & \textbf{UO} & instruction implementation error & Confirmed \\
			                       & \#2988 & \textbf{UO} & instruction implementation error & Confirmed \\
			                       & \#2996 & \textbf{UO} & instruction implementation error & Confirmed \\
			                       & \#2997 & \textbf{UO} & instruction implementation error & Confirmed \\
			                       & \#2999 & \textbf{UO} & instruction implementation error & Confirmed \\
			                       & \#3018 & \textbf{RF} & memory check error               & Confirmed \\
			                       & \#3019 & \textbf{RF} & memory check error               & Confirmed \\
			                       & \#3057 & \textbf{RF} & memory check error               & Confirmed \\
			                       & \#3063 & \textbf{RF} & memory check error               & Confirmed \\
			                       & \#3068 & \textbf{RF} & memory check error               & Confirmed \\
			                       & \#3076 & \textbf{RF} & memory check error               & Confirmed \\
			\hline
		\end{tabular}%
	}
     \vspace{-0.1in}
\end{table}

\noindent \textbf{Bug Reporting.}
For each unique bug we identified, we randomly select binaries and report them to runtime developers, and we further provide root cause analysis to them for aiding bug fixing. For the 33 unique bugs, 25 of them have been confirmed by developers, and 11 of them have been fixed by the time of this writing, as shown in
Table~\ref{table:bugs}.
For those confirmed ones, we will conduct detailed case studies on some representatives in the following \S\ref{sec:evaluation:rq3}

\textbf{Answer to RQ2.} \textit{{\framework} successfully generated more than 167K Wasm binaries that can lead to unexpected behaviors of Wasm runtimes, which were caused by 33 unique bugs. With our timely disclosure, 25 bugs have already been confirmed by runtime developers, and 11 of them have been fixed with our help.}

\subsection{RQ3: Bug Characterization}
\label{sec:evaluation:rq3}

We go a step further to analyze the root cause of each bug manually.
As shown in Table~\ref{table:bugs} illustrates, most of these bugs are caused by instruction implementation errors, while some bugs are due to boundary check and type conversion errors.
We next select three fixed bugs as representative ones for case studies.

\begin{lstlisting}[caption={Wasmtime Type conversion error.}, label={lst:wasmtime}, mathescape=true]
(memory (;0;) 65536 65536)
(data (;0;) (i32.const -79158787) "Bp222N")
\end{lstlisting}
\vspace{-0.05in}

\textbf{Case 1: Wasmtime Type conversion error.}
Listing~\ref{lst:wasmtime} shows a part of the Wasm binary generated by {\framework}, leading to the \#7558 issue of Wasmtime.
Specifically, L1 specifies the maximum space of the linear memory of this binary is 4 GB. However, L2 initializes the memory area starting from the offset of -79158787, which needs to be interpreted as an unsigned 32-bit integer.
Unfortunately, wasmtime incorrectly takes this number as a signed one, leading to a compilation failure as a negative address is invalid.

\begin{lstlisting}[caption={Wamr instruction implementation error.}, label={lst:wamr1}, mathescape=true]
(module
 (func (result v128)
 v128.const i32x4 0x3d52aa71 0xea2f90b2 0xb20cdf3d 0x4d6054bc
 i32.const -7235
 i8x16.shl)
 (export "main" (func 0)))
\end{lstlisting}
\vspace{-0.05in}

\textbf{Case 2: wamr instruction implementation error.}
Listing~\ref{lst:wamr1} illustrates a part of the Wasm binary generated by {\framework} that triggers the \#2690 issue of wamr.
To be specific, the \code{i32.const -7235} at L4 determines how many bits should be left-shifted, and it should be interpreted as the unsigned number 58302.
Unfortunately, wamr incorrectly treats -7235 as a signed number and directly performs modulo with 128. Thus, the following instruction left-shift the \texttt{v128.const} at L3 in different bit.
To this end, though this code snippet can be executed normally, wamr outputs a different output compared with other runtimes.

\begin{lstlisting}[caption={Wamr string check error.}, label={lst:wamr2}, mathescape=true]
(export "\00jCeH" (func 0))
(export "" (func 1))
(export "fj" (func 2))
\end{lstlisting}
\vspace{-0.05in}

\textbf{Case 3: wamr string check error.}
Listing~\ref{lst:wamr2} shows a case that leads to the \#2789 issue of wamr.
The first two export items export \texttt{\textbackslash00jCeH} and an empty string, respectively.
Note that, both of them start with a \texttt{\textbackslash00}.
Instead of using Unicode string to handle import and export names, as required by the Wasm spec, wamr adopts c-style string. Thus wamr mistakenly resolves the first two exports to two empty strings, which causes wamr to fail to compile the bytecode and raise the \textit{duplicate export name} exception.

\textbf{Answer to RQ3.} \textit{Instruction implementation error and memory check error are the top two factors leading to the unexpected behaviors of Wasm runtimes, underlining the importance of performing adequate sanity checks and following the specification.}

\section{Discussion}
\label{sec:discussion}

\noindent \textbf{Ethical Consideration.}
This work aims to identify real-world bugs in Wasm runtimes. We have identified 33 unique bugs, and reported them to runtime developers immediately for responsibility. By the time of this writing, most of them have been confirmed, although some bugs are in the process of fixing. we have repeatedly urged developers to fix them by providing necessary help.

\noindent \textbf{Semantics.}
Although {\framework} tries to generate binaries with rich semantics as much as possible, they might be not as semantically rich as real-world binaries.  
This is mainly due to the fact that some complex control flows within sub-trees of ASTs cannot be fully executed. To reach these deep and nested nodes, it depends on the context provided by the previous ASTs. Although we assign values to local variables in functions, we cannot guatantee that we can restore every contexts.

\noindent \textbf{Scalability.}
Wasm is experiencing fast development, and many new proposals are emerging. {\framework}'s current ability to test the semantics of runtimes might not encompass some new proposals. 
However, we argue that {\framework} can be easily extended to cover the new semantics by incorporating real-world Wasm binaries that exploit new features.
Additionally, we will design more generation and mutation strategies for these new semantics to expand the capabilities of {\framework}.

\section{Related Work}

\noindent \textbf{Wasm binary security.}
Lots of existing work focus on the security side of Wasm binaries~\cite{everything,fuzzm,wasmati,wasp,mswasm,wafl,eosafe,eunomia,risk}.
Daniel et al.~\cite{everything} analyzed the memory issues in Wasm binaries, like stack overflow and heap metadata corruption, and adopted binary instrumentation and AFL to detect such memory bugs~\cite{fuzzm}.
Tools like Wasmati~\cite{wasmati} and Wasp~\cite{wasp} leverage code property graphs and concolic execution, respectively, to identify vulnerabilities in Wasm binaries. MSWasm~\cite{mswasm} addresses Wasm's memory safety issues by introducing semantics related to memory safety for Wasm.

\noindent \textbf{Wasm runtime testing.}
There exist some work in testing Wasm runtimes~\cite{zhang,wasmfuzzer,wadiff,performance}.
Specifically, WADIFF~\cite{wadiff} utilizes symbolic execution on the specification of instructions to generate Wasm binaries to conduct the following differential testing.
Moreover, Zhang et al.~\cite{zhang} develops a pattern-based bug detection framework based on pre-defined domain knowledge. 
Additionally, Jiang et al.~\cite{performance} leverages existing performance testing benchmarks to discover performance issues in server-side Wasm runtimes via differential testing.
However, these approaches still struggle in generating Wasm binaries with rich semantics.

\noindent \textbf{Differential testing.}
McKeeman~\cite{differential} first introduced the concept of differential testing.
Then, it has been extensively employed against various applications, including virtual machines~\cite{classfuzz,classming,javascript}, compilers~\cite{compiler1,compiler2,compiler3}, deep learning frameworks~\cite{tensorscope,audee}, and symbolic execution engines~\cite{symbolengine1}.
This paper borrows the idea of differential testing and applies it to Wasm runtimes.

\section{Conclusion}
This paper presents {\framework}, a novel differential testing framework for Wasm runtimes. {\framework} can generate syntactic-correct and semantic-rich Wasm binaries by disassembling and assembling of real-world Wasm binaries. For further pinpointing the root causes of inconsistencies and locating the bugs, we further propose a root cause identification algorithm that can accurately locate bugs on a function-level or even instruction-level. {\framework} shows great performance via extensive evaluation, and we have uncovered 33 real-world bugs in popular Wasm runtimes.

\balance
\bibliographystyle{ACM-Reference-Format}
\bibliography{WRTester}

\end{document}